\documentclass[12pt]{spieman}  
\usepackage{amsmath,amsfonts,amssymb}
\usepackage{graphicx}
\usepackage{setspace}
\usepackage{tocloft}
\usepackage{lineno}
\usepackage[linesnumbered,lined,boxed,commentsnumbered]{algorithm2e}
\usepackage{algpseudocode}
\title{Parallel single-pixel imaging based on modulation region expansion and overlapping reconstruction}

\author[a]{Yinran Shen}
\author[a,b,*]{Xuri Yao}
\author[c,d,*]{Shijian Li}
\author[a]{Chao Shen}
\author[e]{Yuhao Wang}
\author[a]{Chongwu Shao}
\author[a,*]{Qing Zhao}
\affil[a]{Beijing Institute of Technology, School of Physics, Center for Quantum Technology Research and Key Laboratory of Advanced Optoelectronic Quantum Architecture and Measurements, Beijing, China}
\affil[b]{Beijing Institute of Technology, School of Physics, State Key Laboratory of Chips and Systems for Advanced Light Field Display, Beijing, China}
\affil[c]{Aerospace Information Technology University, College of Photonics and Optical Engineering, Jinan, China}
\affil[d]{Aerospace Information Technology University, College of Photonics and Optical Engineering, Shandong Key Laboratory of Intelligent Photonic Transmission and Sensing, Jinan, China}
\affil[e]{Kunming Institute of Physics, Kunming, China}

\cftpagenumbersoff{figure}
\cftpagenumbersoff{table} 
\begin{document} 
\maketitle

\begin{abstract}
Parallel single-pixel imaging (PSPI) enhances the data acquisition efficiency of single-pixel imaging, but its reconstruction quality depends on a cumbersome and noise-sensitive calibration process. To address this challenge, a PSPI strategy was introduced that leverages modulation region expansion and overlapping reconstruction. This method results in the calibration of modulation of the subregion for each detector, enabling robust operations with undersampled data. It compensates for misalignment via modulation region expansion and overlapping reconstruction, achieving seamless and high-quality imaging that surpasses conventional PSPI in simulations and experiments. Furthermore, this strategy exhibits remarkable robustness, maintaining high imaging quality under extremely nonideal conditions, such as large deflection angles between the array detector and the modulator. This work provides a simple, efficient, and robust framework that simplifies the PSPI workflow and offers broad applicability in high-resolution, high-speed computational imaging.
\end{abstract}


{\noindent \footnotesize\textbf{*}Xuri Yao,  \linkable{yaoxuri@bit.edu.cn} }

{\noindent \footnotesize\textbf{*}Shijian Li,  \linkable{lishijian@aitech.edu.cn} }

{\noindent \footnotesize\textbf{*}Qingzhao,  \linkable{qzhaoyuping@bit.edu.cn} }

\begin{spacing}{2}   

\section{Introduction}
\label{sect:intro}  


As a promising computational imaging technique, single-pixel imaging (SPI) employs a spatial light modulator (SLM) to apply a time-varying structured light field to the target, while a single-pixel detector is used to measure the total light intensity\cite{5,6,28,SPI_ref2,SPI_ref1,APN_ref1}. The image is then computationally reconstructed by correlating the modulations with the intensity measurements. SPI is characterized by a simple workflow, high sensitivity and low data volume and provides an alternative solution for imaging scenarios where array detectors are unavailable or underdeveloped\cite{27}. Considerable potential has been demonstrated for SPI in applications that involve imaging across ultraviolet to terahertz wavebands\cite{ultraviolet_SPI1,infrared_SPI_2,terahertz_SPI_2}, gas detection\cite{11}, light detection and ranging (LiDAR)\cite{lidar1,lidar2}, and photoacoustic imaging\cite{photoacoustic_imaging_1,photoacoustic_imaging_2}.


SPI requires a series of pattern-modulated measurements, making it an inherently time-consuming process. 
The original SPI leverages compressive sensing (CS), which circumvents the sampling limitations imposed by the Shannon-Nyquist theorem, to reduce the number of measurements\cite{7,16,26}. Nevertheless, a trade-off remains between temporal efficiency and spatial resolution, precluding the simultaneous achievement of high frame rates and high spatial resolution\cite{limination_of_SPI1}. 
To mitigate this challenge, high-speed modulation techniques\cite{17,18} and low-sampling-rate reconstruction methods\cite{29,deep_learning_low_sampling_rate_1} have been employed in SPI systems. With respect to high-speed modulation approaches, devices such as high-speed light-emitting diodes (LEDs) can achieve MHz-level modulation rates but face challenges such as high system complexity, inflexible modulation schemes, and limited spatial resolution\cite{SLM_limination_1}. Although methods such as orthogonal basis ordering and deep learning (DL) can effectively reduce the number of measurements, they rely heavily on prior knowledge or large-scale datasets\cite{27}.


Parallel SPI (PSPI) provides an effective means of balancing the frame rate and spatial resolution. In this approach, the conventional single-pixel detector is replaced with a few-pixel array detector, and the SLM is divided into multiple modulation subregions. Each subregion is mapped to a corresponding pixel of the array detector, thereby forming an array of “single-pixel cameras” that operate in parallel\cite{2,4,parallel_SPI_ref1,19,20,31,32,parallel_by_Talbot_effect,parallel_frequency_division_multiplexed,parallelSPI_ref_1,parallel_work_IR1}. Since there are only a few pixels (typically $4\times 4$ pixels) in each subregion, the total number of required modulations is significantly reduced, which can result in high-speed imaging functionality. Although an array detector is used in PSPI, the pixel resolution of the reconstructed image exceeds that of the detector array, offering a form of “pixel superresolution imaging” compared with conventional focal-plane imaging methods. As such, this technique is particularly well suited for imaging scenarios involving detectors with a limited number of pixels, such as InGaAs or HgCdTe array detectors. 
For instance, a PSPI system in a previous study achieved an imaging resolution of ${1280\times1024}$ pixels at 20 fps using a mid-wave infrared camera with a native resolution of only ${320\times256}$ pixels\cite{3}.
Significant potential has also been demonstrated for PSPI in the fields of 3D imaging\cite{24,parallel_3D_imaging1,Parallel_SPIand3D_1,Parallel_SPIand3D_2} and large field-of-view (FOV) imaging\cite{35}.

However, the precise alignment between modulation subregions and detector pixels poses a challenge in PSPI. 
The pixel sizes of detectors and SLM differ, and particularly in PSPI implemented using a digital micromirror device (DMD)—which is a common choice because of its high modulation frame rate—its tendency for reflective modulation makes precise alignment even more challenging.
Therefore, calibration between each modulation unit and the detector pixel is essential, and the mapping relationship between them is defined as the transfer function matrix (TFM).
The TFM is crucial for image restoration, yet it is often obtained through point scanning\cite{2,4}, which is an inefficient process with a low signal-to-noise ratio (SNR). 
Although methods such as CS-based methods have been proposed for efficiently obtaining the TFM\cite{1} and thereby overcoming the abovementioned limitations, achieving high accuracy still requires full sampling, which is a time-consuming and complex process.
In this study, we propose a PSPI strategy that employs modulation region expansion and overlapping reconstruction (EOPSPI).
By directly determining the modulation subregion for each detector from undersampled calibration data, our method circumvents the TFM and its associated drawbacks, which include estimation noise and the requirement for full sampling. To correct the misalignment between the array detector and the DMD, we combine a modulation region expansion scheme with an overlapping reconstruction approach, yielding seamless, high-fidelity reconstruction. Results of extensive simulations and experiments reveal that compared with conventional calibration techniques, the proposed method consistently delivers superior image quality.
Furthermore, we demonstrate that the proposed method maintains satisfactory imaging quality even under large deflection angles between the array detector and the DMD, highlighting its adaptability to more challenging environments. 
This work provides a simple workflow and improves the quality of PSPI. In addition, the implementation conditions are relaxed in this method and it can be readily adapted to other modalities that require pixel-level alignment, including panoramic imaging\cite{panoramic_imaging_1,panoramic_imaging_2} or adaptive optics wavefront sensing\cite{SHWS_1,SHWS_2}.

\section{Principle}
\label{sect:Principle and simulation}
\subsection{Principle of PSPI}
\label{sect:Principle of parallel single-pixel imaging}

\begin{figure}[t]
\begin{center}
\begin{tabular}{c}
\includegraphics[height= 16.2 cm]{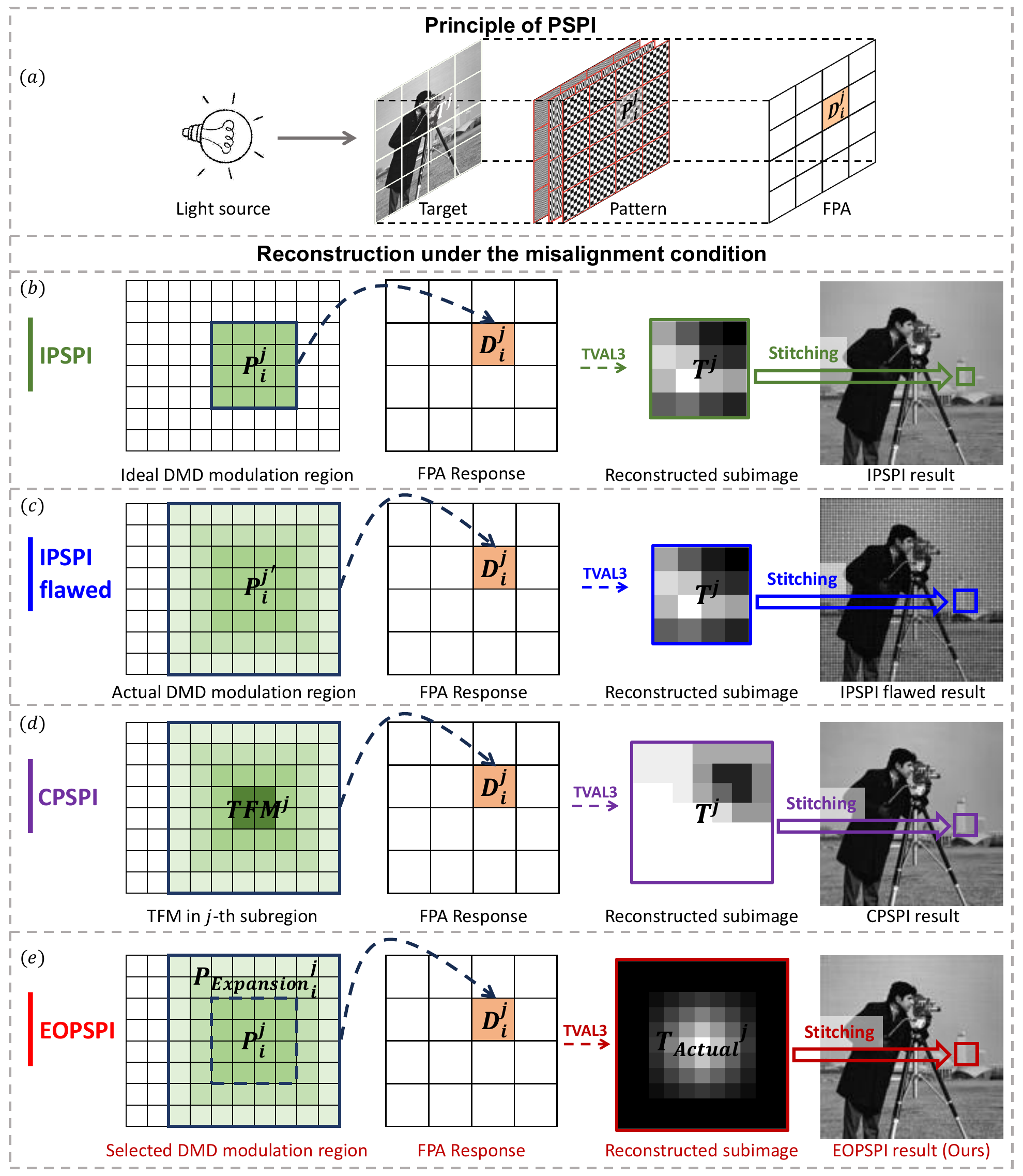}
\end{tabular}
\end{center}
\caption 
{ \label{fig:pinciple_of_the_proposed_method}
Comparison of different PSPI methods.
(a) Principle of the PSPI method.
(b) Schematic of the mapping relationship and the reconstructed results of the IPSPI method. (c) Flawed reconstruction results obtained in IPSPI. (d) Schematic of the mapping relationship and the reconstructed results of the CPSPI method. (e) Schematic of the mapping relationship and the reconstructed results of the EOPSPI method.} 
\end{figure}

In PSPI, the modulation pattern can be partitioned into a grid of $m$ adjacent subpatterns, each modulating its corresponding subimage in parallel. As shown in Figure~\ref{fig:pinciple_of_the_proposed_method}(a), the subimage $T^{j}$ corresponds to the region of the target modulated by the subpattern $P^{j}$. Each single-pixel detector subsequently independently records the intensity values of the correlations between its corresponding subpatterns and the target. The intensity measurement for the $i$-th sampling is expressed as follows:
\begin{equation}
\label{eq:parallel_imaging_theoretical_eqa}
D_{i}^{j} = \sum_{x}\sum_{y}P_{i}^{j}(x,y)T^{j}(x,y),
\end{equation}
where $(x,y)$ denotes the Cartesian coordinate, $P_{i}^{j}$ represents the $j$-th subpattern in the $i$-th modulation pattern, $T^{j}$ is the $j$-th subregion of the target, $D_{i}^{j}$ is the light intensity measurements received by a single-pixel detector corresponding to $P_{i}^{j}$, $i$ is the measurement number ($i = 1,2,..,N$), and $j$ represents the number of subregions ($j = 1,2,..,m$).
Then, the target can be reconstructed by correlating the modulation patterns with their corresponding intensity measurements. 
In this paper, a CS method of TVAL3\cite{34} is used for recovering the subimage. 


For the ideal PSPI (IPSPI), Figure~\ref{fig:pinciple_of_the_proposed_method}(b) depicts this mapping relationship, where modulation subregions map precisely to the detection units of the FPA.
The subimages are then reconstructed and stitched together. 
Owing to misalignments, crosstalk, and optical aberrations, in real experiments, the modulation subregions received by the detection units are mismatched with their ideal counterparts.
As a result, this precise subdivision of the modulation patterns is problematic. 
For example, as shown in Figure~\ref{fig:pinciple_of_the_proposed_method}(c), the actual modulation subregion is larger than the ideal subregion. Consequently, the reconstruction introduces obvious artifacts and seams.

As shown in Figure~\ref{fig:pinciple_of_the_proposed_method}(d), a common approach for addressing practical limitations is to calibrate the modulation matrix via the TFM, which describes the mapping between the FPA and the DMD. In calibration-based PSPI (CPSPI), the intensity values received by the $j$-th detector are expressed as follows:
\begin{equation}
\label{eq:light_collected_TFM_eqa}
{D_{CPSPI}}_{i}^{j} = \sum_{x}\sum_{y}P_{i}^{j}(x,y)TFM^{j}(x,y)T^{j}(x,y).
\end{equation}
Figure~\ref{fig:pinciple_of_the_proposed_method}(d) shows the seamless reconstruction obtained by TVAL3.
However, the reconstruction quality is critically dependent on the accuracy of the calibration. In practice, TFMs exhibit complex distributions and lack translation invariance, necessitating independent measurements for each detector. To obtain a TFM with high accuracy, calibration is typically performed at full sampling\cite{1,2,4}. This pursuit of accuracy nevertheless entails a substantial experimental burden.

\subsection{Principle of the proposed EOPSPI}
\label{sect:The principle of the proposed method}
As discussed, this misalignment causes the modulation subpattern to vary in both size and permutation relative to the ideal subpattern. The impact of permutation on reconstruction is analyzed in one study\cite{parallel_SPI_ref1}. Therefore, the modulation subregion received by each detector must be redetermined.
As in conventional calibration methods, a uniformly illuminated white Lambertian board is employed as the target, positioned to cover the FOV of the DMD. This target is modulated by the modulation patterns, and then the intensity values are recorded in parallel by the array detector. For each detector, an image is reconstructed from its undersampled measurements using CS. In the reconstructed image for a given detector, the region where the pixel intensity significantly exceeds the background corresponds to the area of the DMD that effectively modulates light onto that detector. This region is defined as the actual modulation subregion for that detector by applying an intensity threshold.
Technical details are provided in Supplementary Materials Sec. 1.
As illustrated in Figure~\ref{fig:pinciple_of_the_proposed_method}(e), the EOPSPI approach incorporates the effects of the imaging system on modulation subpatterns into both the modulation and reconstruction processes. Specifically, EOPSPI introduces an expanded margin around each modulation region $P_{i}^{j}$. 
The expanded margin region ${P_{Expansion}}_{i}^{j}$, shown in Figure~\ref{fig:pinciple_of_the_proposed_method}(e), prevents truncation of the modulation subpattern in each subregion, which ensures complete modulation information acquisition.
This actual modulation region ${P_{Actual}}_{i}^{j}$, which is an expansion of the modulation region $P_{i}^{j}$, is then defined as the modulation subpattern.
The intensity measurement ${D_{Actual}}_{i}^{j}$ for the $j$-th single-pixel detector is subsequently expressed as follows:
\begin{align}
\label{eq:light_collected_proposed_method_eqa}
{D_{Actual}}_{i}^{j} &= \sum_{x}\sum_{y}{P_{Actual}}_{i}^{j}(x, y){T_{Actual}}^{j}(x, y),
\end{align}
where ${P_{Actual}}_{i}^{j}$ is the actual modulation subpattern in the $i$-th modulation pattern that corresponds to the $j$-th single-pixel detector in the experiments, ${T_{Actual}}^{j}$ is the region of the target modulated by ${P_{Actual}}_{i}^{j}$, ${D_{Actual}}_{i}^{j}$ is the intensity measurement captured by the single-pixel detector, and $j$ is the index of the subregion with $j = 1,2,..,m$. 

\begin{figure}[htbp]
\begin{center}
\begin{tabular}{c}
\includegraphics[height= 6.9 cm]{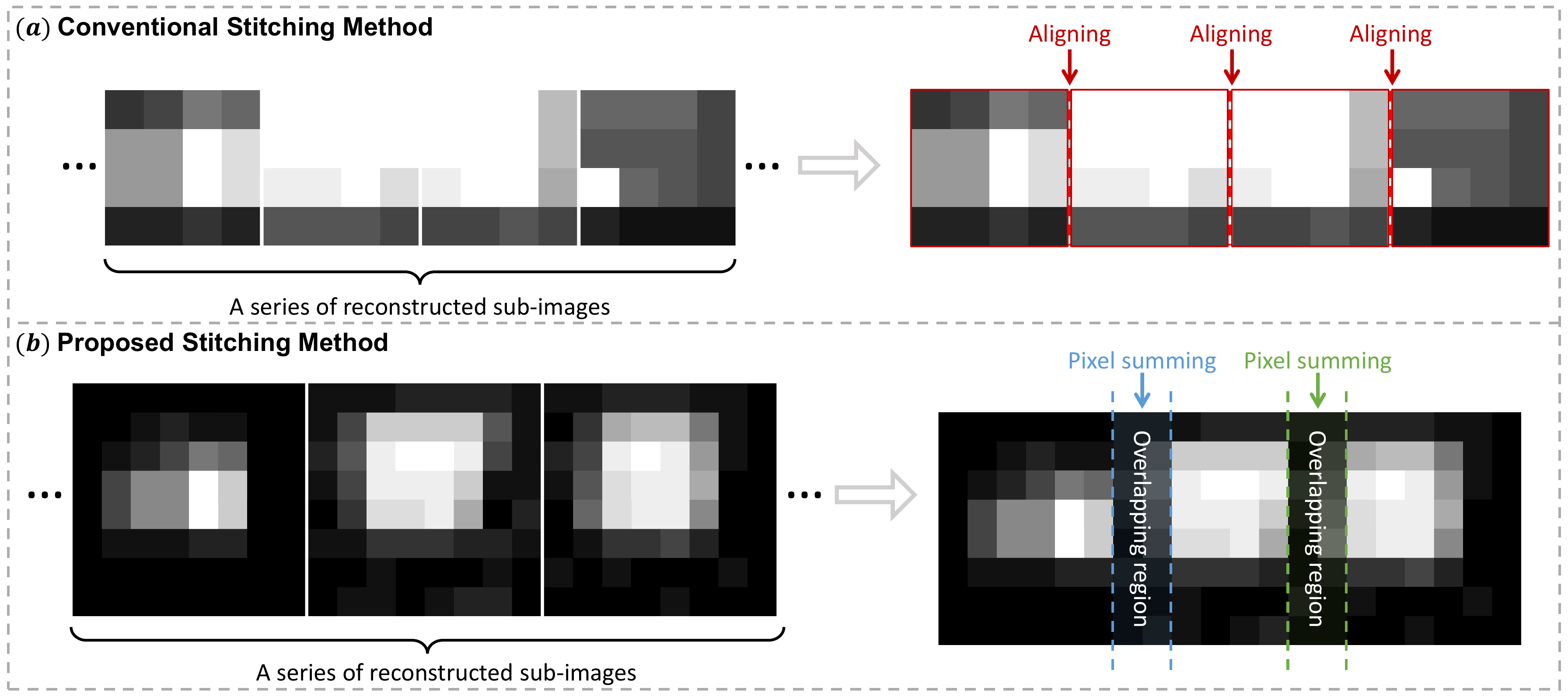}
\end{tabular}
\end{center}
\caption 
{ \label{fig:stitching_method}
Comparison of different stitching methods. (a) Conventional stitching method used in IPSPI. (b) Stitching method used in our study.} 
\end{figure}

The subimages are reconstructed sequentially from the first detector to the $m$-th detector and then stitched together. Unlike the stitching method of simply aligning subreconstructions in IPSPI, Figure~\ref{fig:stitching_method} demonstrates a comparison between the stitching methods used in IPSPI and our approach. The proposed stitching method results in the generation of high-resolution images by summing the pixels in the overlapping region. The seamless stitched image obtained by the proposed stitching method is shown in Figure~\ref{fig:pinciple_of_the_proposed_method}(e).
Our method not only eliminates the seams and artifacts of the IPSPI approach but also delivers visual quality comparable to that of the CPSPI method.

\section{Experiments}
\label{sect:Experiments}

\begin{figure}[htbp]
\begin{center}
\begin{tabular}{c}
\includegraphics[height= 6.7 cm]{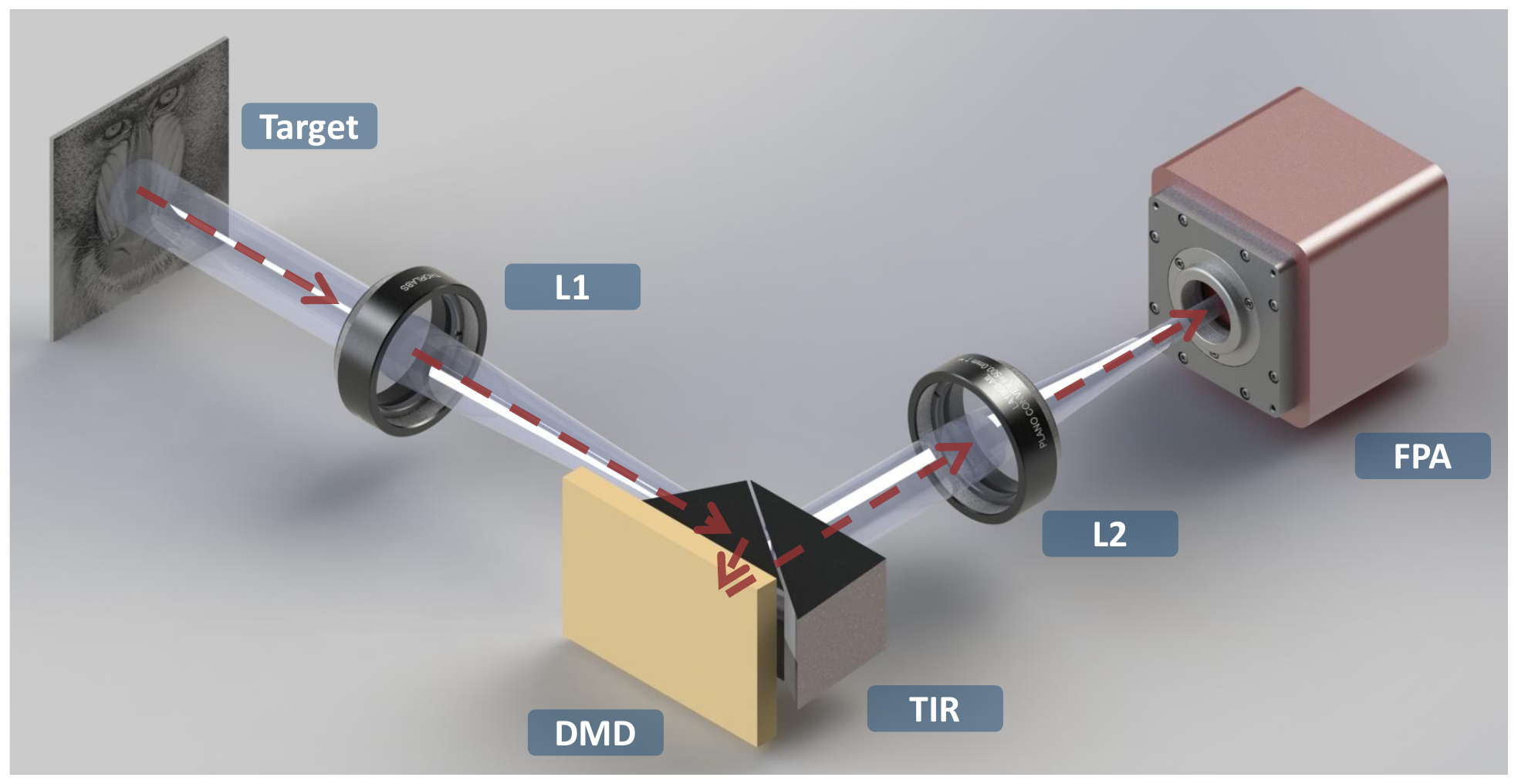}
\end{tabular}
\end{center}
\caption 
{ \label{fig:experiment_setup}
Schematic of the experimental setup.} 
\end{figure}

To validate the effectiveness of our method, as shown in Figure~\ref{fig:experiment_setup}, a structured detection experimental setup was employed. 
The target is focused onto the modulation plane of the DMD via a lens L1 ($f = 80$ mm) and a total-internal reflection (TIR) prism. The DMD (Vialux, DLP6500) is then used to introduce time-varying spatial modulation to the target.
The modulated light field is subsequently relayed to the imaging plane of the FPA (DAHENG Optics, MER-041-436U3M) by a lens L2. 
In the experiment, the FPA and the DMD are synchronized. L2 approximately maps 1 pixel of the FPA to $4\times 4$ pixels of the DMD.
The modulation patterns are generated by replicating $8\times 8$ pixel Hadamard subpatterns to the modulation region of the DMD. Notably, the modulation subpattern is marginally oversized relative to the actual DMD-FPA mapping to suppress crosstalk caused by modulation region expansion. 
We present the results for an expansion factor of $k = 2$, which indicates the pixel width by which the ideal modulation region margin is expanded.

\begin{figure}[t]
\begin{center}
\begin{tabular}{c}
\includegraphics[height= 10.8 cm]{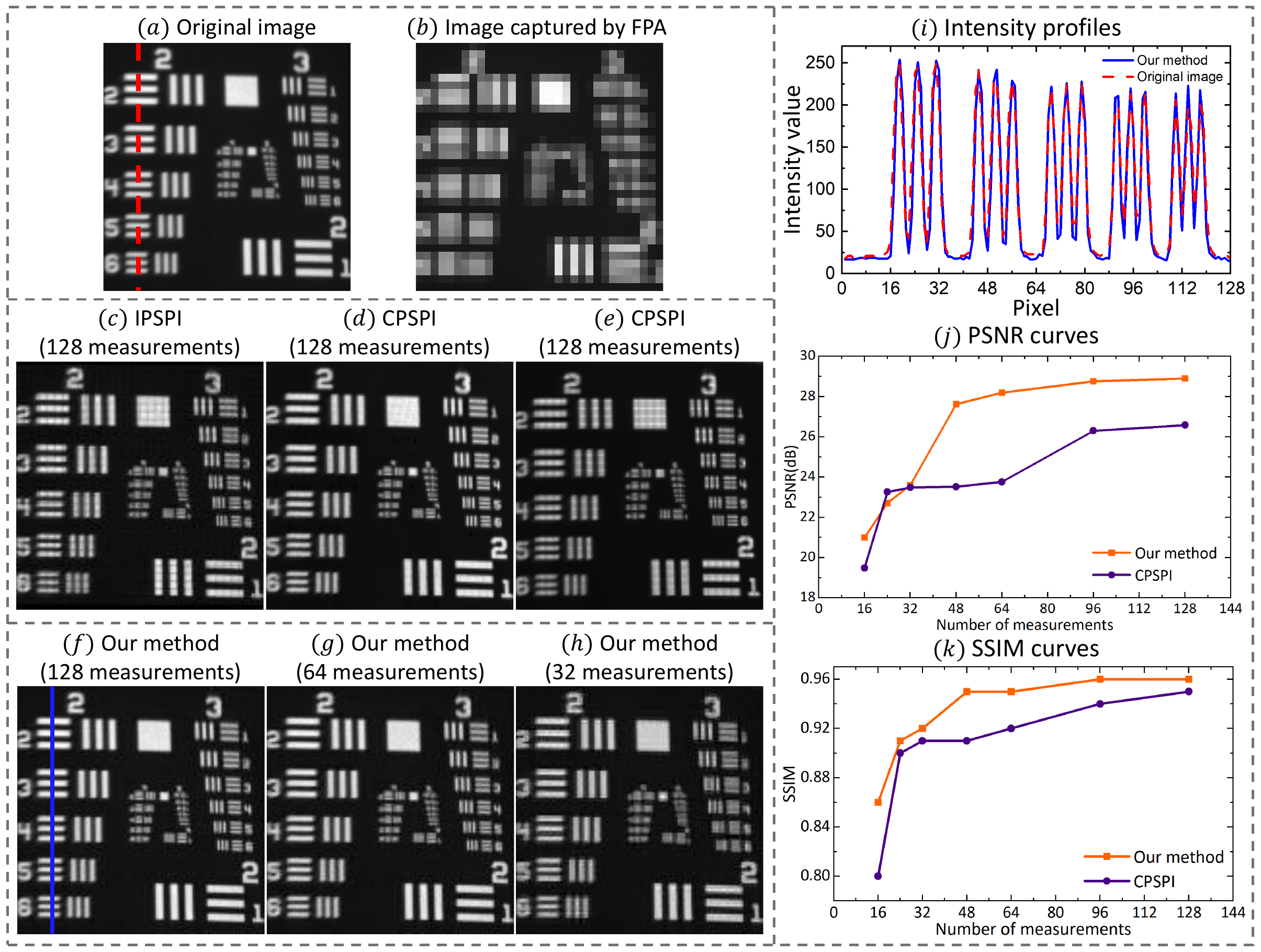}
\end{tabular}
\end{center}
\caption 
{ \label{fig:usaf_1951_nontilt_1}
Imaging results and quantitative analysis of a 1951 USAF resolution test chart. (a) Original image. (b) Low-resolution image captured by the FPA. (c) Reconstructed image by the IPSPI method with 128 measurements. (d)–(e) Reconstructed images by the CPSPI method (under full-sampling and a 16\% sampling rate during calibration) with 128 measurements. (f)–(h) Reconstructed images by the proposed method using 128, 64 and 32 measurements, respectively. (i) Cross-sectional intensity profiles of the red dashed line and blue line shown in (a) and (f). (j)–(k) Quantitative comparison of PSNR and SSIM for the proposed method and CPSPI. } 
\end{figure}

To investigate the superresolution capability of EOPSPI, imaging experiments were performed using a USAF 1951 resolution target.
Figure~\ref{fig:usaf_1951_nontilt_1}(a) and \ref{fig:usaf_1951_nontilt_1}(b) present the original image and the low-resolution reference image (of $32\times 32$ pixels captured by the FPA), respectively. Limited by insufficient spatial resolution, this image fails to reveal key structural features. 
The PSPI method is then used to achieve $\times 4$ superresolution imaging. This requires the modulation patterns of $128\times 128$ pixels to reconstruct images at the matching resolution.
As shown in Figure~\ref{fig:usaf_1951_nontilt_1}(c), the misalignment between the DMD and the FPA results in visible seams in the IPSPI reconstruction.
The reconstructed results of the CPSPI obtained with 128 measurements when the TFM is obtained with full sampling and a 16\% sampling rate during calibration are shown in Figure~\ref{fig:usaf_1951_nontilt_1}(d) and \ref{fig:usaf_1951_nontilt_1}(e), respectively.
The reconstruction quality decreases as the sampling rate of the calibration decreases, whereas achieving high accuracy through full sampling increases the experimental time and complexity. However, even with full sampling, environmental noise introduces errors in the calibration process, leading to noticeable artifacts and stripes (a detailed description is given in Sec. 2 of the Supplementary Materials). 
Consistent with the calibration condition (16\% sampling rate) in Figure~\ref{fig:usaf_1951_nontilt_1}(e), Figure~\ref{fig:usaf_1951_nontilt_1}(f)–~\ref{fig:usaf_1951_nontilt_1}(h) show the reconstructed images of the EOPSPI strategy with 128, 64, and 32 measurements, respectively. Experimental results for other expansion factors and different measurements are given in the Supplementary Materials Secs. 3 and 4.
Compared with the images reconstructed by IPSPI and CPSPI, seamless reconstructions were achieved by our method. Notably, even with as few as 32 measurements, the reconstructed image still maintains visually acceptable quality. 
To evaluate the reconstruction performance, as shown in Figure~\ref{fig:usaf_1951_nontilt_1}(i), the cross-sectional intensity profiles of both the reconstructed image obtained through 128 measurements [as shown in Figure~\ref{fig:usaf_1951_nontilt_1}(f)] and the original image [as shown in Figure~\ref{fig:usaf_1951_nontilt_1}(a)] were analyzed. The red dashed line indicates the position of the 19th column in the original image, and the blue line indicates the position of the same column in the reconstructed image.
The results revealed that the reconstructed intensity profile obtained via the EOPSPI method is in good agreement with the original image.

To quantitatively evaluate the reconstruction performance under the different measurements, the peak signal-to-noise ratio (PSNR) and the structure similarity index (SSIM) were used as assessment metrics (definitions are given in Sec. 5 of the Supplementary Materials).
As plotted in Figure~\ref{fig:usaf_1951_nontilt_1}(j) and \ref{fig:usaf_1951_nontilt_1}(k), for the two methods, both the PSNR and the SSIM increase with increasing measurements, with our method consistently outperforming the CPSPI across the entire sampling range.


\begin{figure}[htbp]
\begin{center}
\begin{tabular}{c}
\includegraphics[height= 9.6 cm]{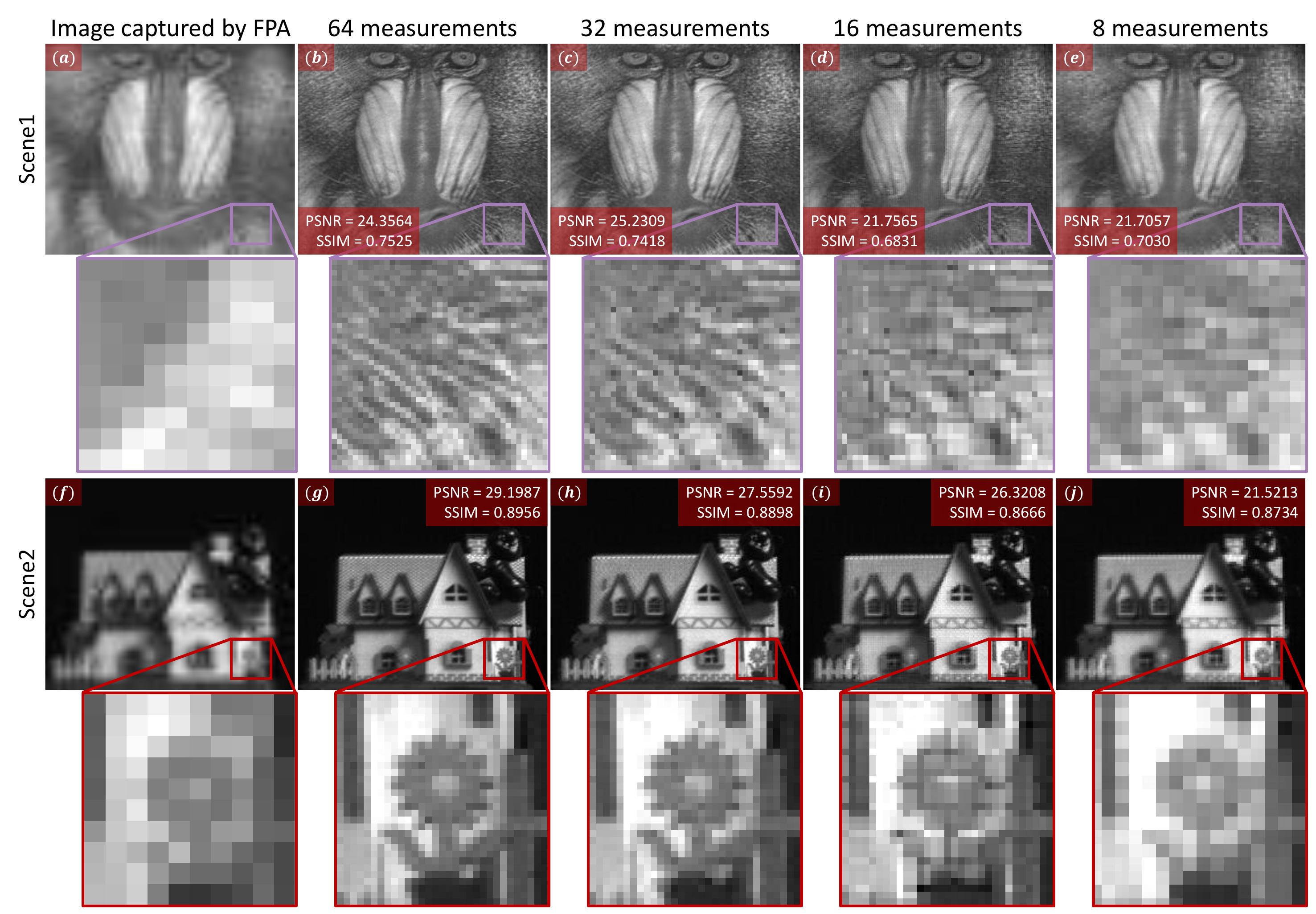}
\end{tabular}
\end{center}
\caption 
{ \label{fig:complex_scene_nontilt_4}
Experimental results for complex scenes with 64, 32, 16, and 8 measurements. } 
\end{figure} 

As shown in Figure~\ref{fig:complex_scene_nontilt_4}, we next evaluated the imaging performance of EOPSPI in more complex scenarios.
Two different targets are imaged: the standard test image “Baboon” (printed on A4 paper) and a 3D scene composed of a toy house and a black toy cat. 
The low-resolution images are captured by the $54\times 64$-pixel FPA. 
Figure~\ref{fig:complex_scene_nontilt_4}(b)–~\ref{fig:complex_scene_nontilt_4}(e) and ~\ref{fig:complex_scene_nontilt_4}(g)–~\ref{fig:complex_scene_nontilt_4}(j) demonstrate the reconstructed images of $216\times 256$ pixels with the different measurements (64, 32, 16, and 8). The experimental results reveal that the reconstructed images with different measurements exhibit no obvious global difference, but significant differences can be observed in the enlarged details. Moreover, even under the condition of extreme undersampling (as few as 8 measurements), the reconstructed images retain visually discernible details. These results collectively reveal the robustness of the EOPSPI strategy to image complex scenes and its ability to image scenes with few measurements.

\begin{figure}[b]
\begin{center}
\begin{tabular}{c}
\includegraphics[height= 6.5 cm]{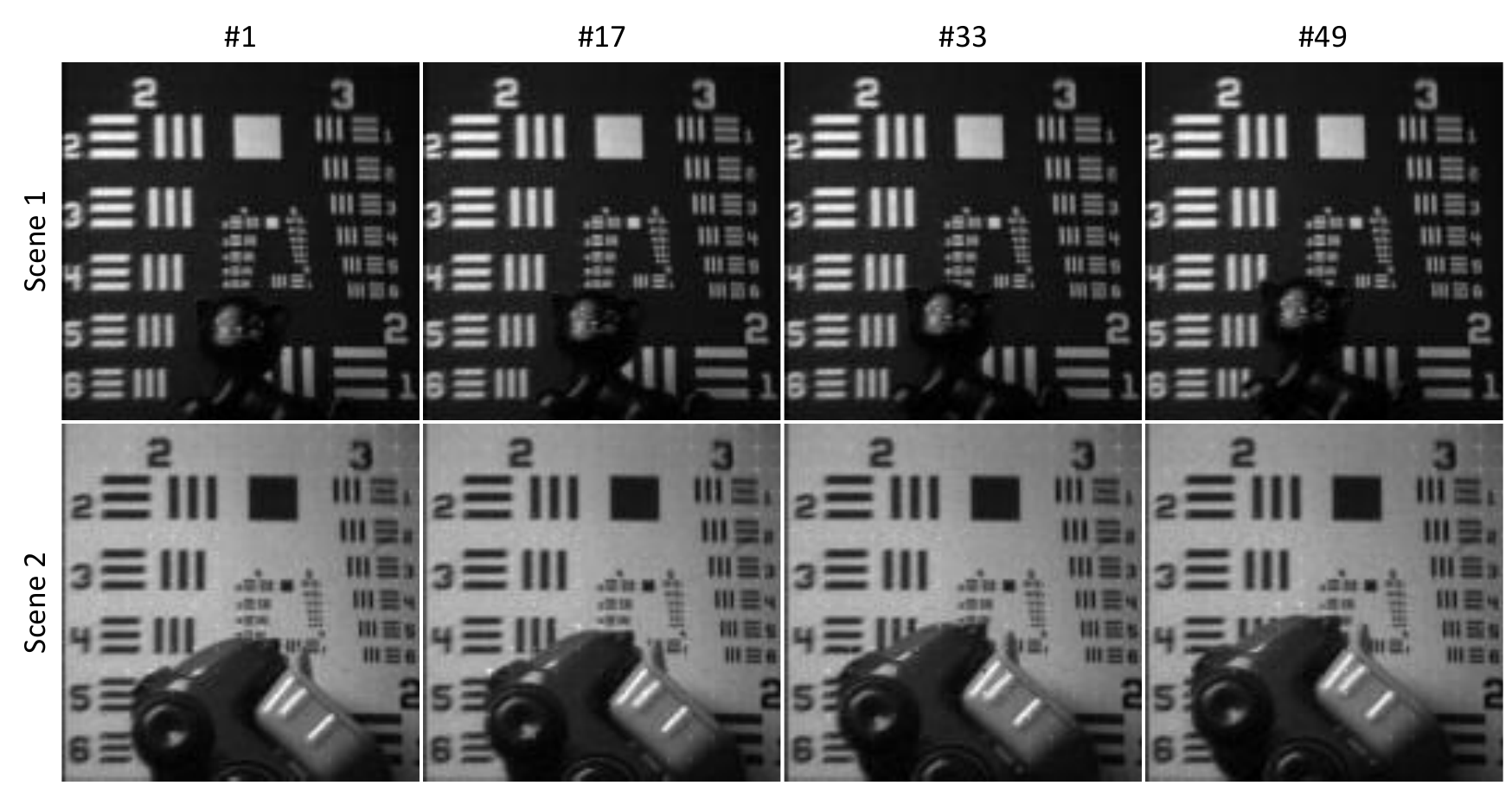}
\end{tabular}
\end{center}
\caption 
{ \label{fig:video_imaging_result}
Selected frames from the dynamic video. } 
\end{figure} 

We further evaluated the performance of the EOPSPI in dynamic scenarios. Figure~\ref{fig:video_imaging_result} presents the imaging results for the two moving scenes.
The background of both scene 1 and scene 2 was the USAF 1951 resolution test target. The motion targets (a black cat in scene 1 and a toy car in scene 2) were driven by a motor. 
In this experiment, the exposure time of the FPA was set as 4 ms. TThe resolution of the reconstructed images was $128\times 128$ pixels, and they were all obtained with 64 measurements. Together, these parameters enable an effective imaging rate of 4 fps.
The complete videos can be found in Supplementary Movie 1. 
The reconstructed images reveal that despite noticeable displacement between frames, their features can be clearly reconstructed without motion artifacts. This result confirms the ability of the EOPSPI method to image dynamic scenes.

Beyond the experimental results presented above, we further evaluated the generalizability of EOPSPI through extensive simulations. Reconstruction performance under various optical parameters, together with detailed comparisons with CPSPI, are provided in Secs. 6–8 of the Supplementary Materials.

\section{Robustness evaluation}
\label{sect:Robustness Evaluation}



\begin{figure}[htbp]
\begin{center}
\begin{tabular}{c}
\includegraphics[height= 9.8 cm]{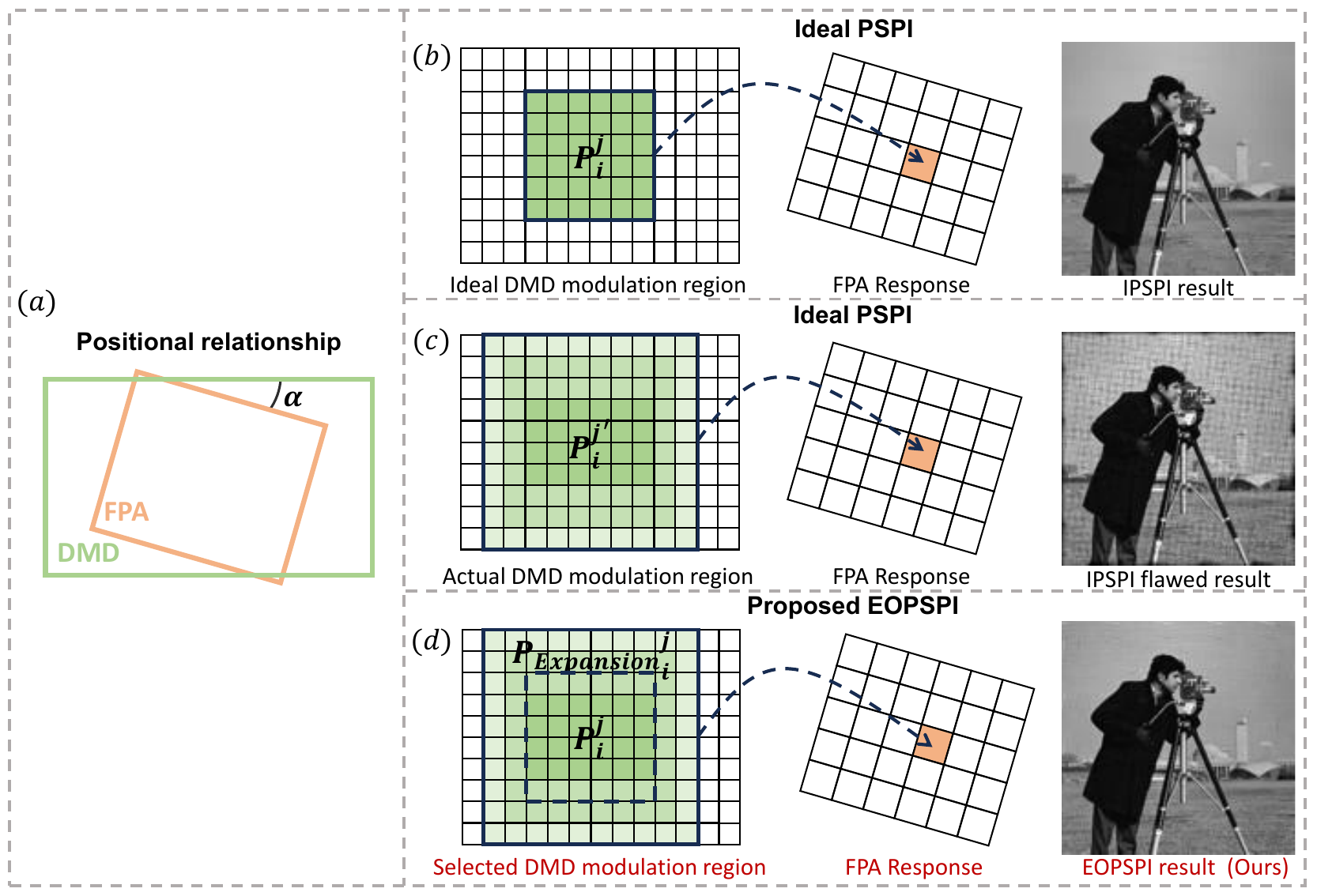}
\end{tabular}
\end{center}
\caption 
{ \label{fig:completely_misalignment_3}
Robustness under extreme conditions. (a) Position relationship between the FPA and the DMD. (b) Schematic of the mapping relationship and the reconstructed results obtained by the IPSPI method. (c) Flawed reconstruction by IPSPI. (d) Schematic of the mapping relationship and the reconstructed results obtained by the EOPSPI method.} 
\end{figure}

To assess the robustness of the EOPSPI against severe system imperfections, we considered an extreme nonideal scenario involving a large angular deflection between the FPA and the DMD, as shown in Figure~\ref{fig:completely_misalignment_3}.
The positional relationship between the FPA and the DMD is quantitatively represented in Figure~\ref{fig:completely_misalignment_3}(a), in which the red rectangular line denotes the imaging plane of the FPA and the green rectangular line denotes the modulation plane of the DMD. The deflection angle between them is denoted by $\alpha$.
The ideal high-resolution reconstruction in this case is shown in Figure~\ref{fig:completely_misalignment_3}(b).
However, as shown in Figure~\ref{fig:completely_misalignment_3}(c), the actual modulation subregions deviate from their ideal regions, leading to obvious seams in the reconstruction.
The high-resolution image reconstructed by the EOPSPI method is shown in Figure~\ref{fig:completely_misalignment_3}(d).
These results confirm that EOPSPI provides robust imaging even under significant nonideal conditions.


To experimentally validate the performance of the EOPSPI under such extreme conditions, measurements were performed using the structured detection setup shown in Figure~\ref{fig:experiment_setup}.
In this experiment, the deflection angle was set to 12° and the expansion factor was set to 2.
This is a parameter that optimally trade-off between reconstruction quality and imaging efficiency.

\begin{figure}[b]
\begin{center}
\begin{tabular}{c}
\includegraphics[height= 5.9 cm]{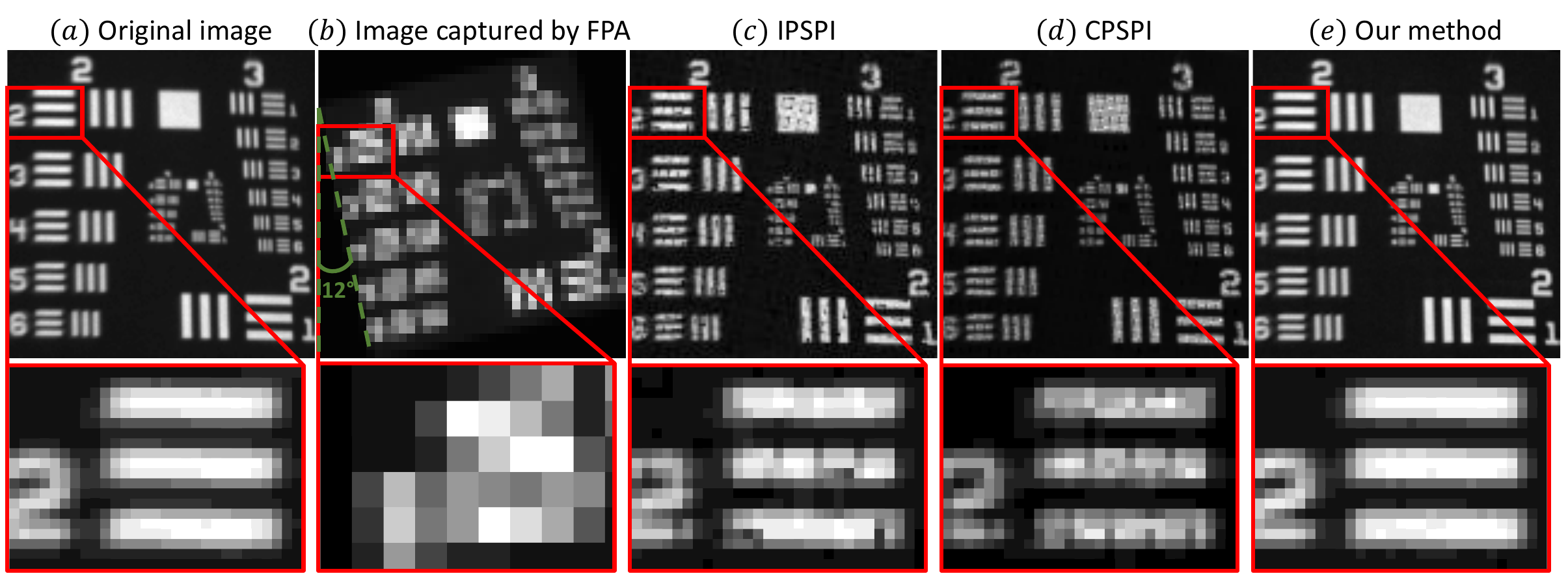}
\end{tabular}
\end{center}
\caption 
{ \label{fig:USAF_1951_tilt2}
Experimental results of the USAF 1951 resolution test chart. (a) Original image. (b) Image captured by the FPA. (c)–(e) Images reconstructed via IPSPI, CPSPI, and our method, respectively.} 
\end{figure} 

Figure~\ref{fig:USAF_1951_tilt2} compares the reconstruction performances of the different methods under significant FPA-DMD deflection.
A low-resolution image of $38\times 38$ pixels is captured by the FPA.
As shown in Figure~\ref{fig:USAF_1951_tilt2}(c), we first perform the reconstruction ($128\times 128$ pixels) using the IPSPI method with 128 measurements.
Owing to the mismatch between the predetermined subpattern and the actual modulation subpattern, the reconstructed image exhibits obvious seams and defects.
Owing to the calibration errors of the TFM under conditions of undersampling, the CPSPI method produces an image with more artifacts and stripes than the IPSPI method.
We quantify this effect by analyzing the impact of varying TFM error levels on reconstruction quality in Sec. 9 of the Supplementary Materials.
Under the condition of 16\% sampling rate, Figure~\ref{fig:USAF_1951_tilt2}(e) presents the seamless reconstruction obtained by the EOPSPI method with 128 measurements, surpassing both IPSPI and CPSPI in fidelity.
The results for other expansion numbers and different measurements can be found in Secs. 10 and 11 of the Supplementary Materials. 
To validate the imaging ability of the EOPSPI method for complex scenes, we performed the experiments detailed in Supplementary Materials Sec. 12.
Furthermore, the robustness of the proposed method was assessed across diverse optical parameters, and a comprehensive comparison with the CPSPI method was performed (details are given in Secs. 13-15 of the Supplementary Materials).

\section{Conclusion}
\label{sect:Conclusion}
In summary, we proposed a PSPI strategy that overcomes the limitations related to the cumbersome, time-consuming, and noise-sensitive system calibration required in conventional PSPI.
Our approach involves the determination of the modulation subregion for each detector, enabling robust calibration using undersampled data (16\% sampling rate) and thereby eliminating the need for calibration at full sampling in conventional PSPI methods.
To compensate for the misalignment between the FPA and the DMD, modulation region expansion is combined with an overlapping stitching algorithm, enabling seamless and high-fidelity imaging.
Results of extensive simulations and experiments confirm that our framework consistently outperforms the conventional PSPI in reconstruction quality.
Crucially, the proposed strategy exhibits exceptional robustness under extreme nonideal conditions, such as large deflection angles between the FPA and the DMD, for which conventional methods fail with severe artifacts seen on the images. 
By transforming the calibration workflow into an efficient and robust workflow, the proposed method results in the simplification of the system setup, acceleration of data acquisition, and reduction in operational complexity. This work establishes a pathway for high-speed and high-resolution computational imaging, with promising applications in fields such as biomedical imaging and industrial inspection.

\section* {Disclosures}
The authors declare no conflict of interest.

\section* {Code and Data Availability}
The data that support the findings of this study are not publicly available at this time but are available from the corresponding author upon reasonable request.

\subsection* {Acknowledgments}
This work was supported by the National Natural Science Foundation of China (Grant No. 12504337), the Beijing Institute of Technology Research Fund Program for Young Scholars (Grant No. 20212012), and Fundamental and Interdisciplinary Disciplines Breakthrough Plan of the Ministry of Education of China (JYB2025XDXM115).


\bibliography{ref}   
\bibliographystyle{spiejour}   



\vspace{1ex}
\noindent Biographies of the authors are not available.


\end{spacing}
\end{document}